\documentclass[fleqn,usenatbib,a4paper]{mnras}

\usepackage{newtxtext,newtxmath}

\usepackage[T1]{fontenc}
\usepackage{ae,aecompl}
\usepackage{times}

\usepackage{etoolbox}
\makeatletter
\patchcmd\@combinedblfloats{\box\@outputbox}{\unvbox\@outputbox}{}{\errmessage{\noexpand patch failed}}
\makeatother

\usepackage{graphicx}	
\usepackage{amsmath}	
\usepackage{amssymb}	

\newcommand{\kms}{\,\mathrm{km\,s}^{-1}}
\newcommand{\kpc}{\,\mathrm{kpc}}
\newcommand{\masyr}{\,\mathrm{mas\,yr}^{-1}}
\newcommand{\percent}{\,\mathrm{per\:cent}}

\title[Tidal disruption of Crater II]{Tidal disruption of dwarf spheroidal galaxies: the strange case of Crater II}

\author[J. L. Sanders et al.]{
Jason L. Sanders,$^{1}$\thanks{E-mail: jls,nwe@ast.cam.ac.uk (JLS, NWE)}
N. W. Evans,$^{1}$
and W. Dehnen$^2$.
\\
$^{1}$Institute of Astronomy, Cambridge, CB3 0HA, UK\\
$^{2}$Department of Physics and Astronomy, University Rd, Leicester, LE1 7RH, UK
}

\date{Accepted XXX. Received YYY; in original form ZZZ}

\pubyear{2017}

\begin{document}
\label{firstpage}
\pagerange{\pageref{firstpage}--\pageref{lastpage}}
\maketitle

\begin{abstract}
Dwarf spheroidal galaxies of the Local Group obey a relationship between the line-of-sight velocity dispersion and half-light radius, although there are a number of dwarfs that lie beneath this relation with suppressed velocity dispersion. The most discrepant of these (in the Milky Way) is the `feeble giant' Crater II. Using analytic arguments supported by controlled numerical simulations of tidally-stripped flattened two-component dwarf galaxies, we investigate interpretations of Crater II within standard galaxy formation theory. Heavy tidal disruption is necessary to explain the velocity-dispersion suppression which is plausible if the proper motion of Crater II is $(\mu_{\alpha*},\mu_\delta)=(-0.21\pm0.09,-0.24\pm0.09)\mathrm{mas\,yr}^{-1}$.
Furthermore, we demonstrate that the velocity dispersion of tidally-disrupted systems is solely a function of the total mass loss even for weakly-embedded and flattened systems. The half-light radius evolution depends more sensitively on orbital phase and the properties of the dark matter profile. 
The half-light radius of weakly-embedded cusped systems rapidly decreases producing some tension with the Crater II observations. This tension is alleviated by cored dark matter profiles, in which the half-light radius can grow after tidal disruption. The evolution of flattened galaxies is characterised by two competing effects: tidal shocking makes the central regions rounder whilst tidal distortion produces a prolate tidally-locked outer envelope. After $\sim70\percent$ of the central mass is lost, tidal distortion becomes the dominant effect and the shape of the central regions of the galaxy tends to a universal prolate shape irrespective of the initial shape. 
\end{abstract}




\section{Introduction}
The dwarf spheroidal galaxies (dSphs) of the Local Group lie at the extreme low mass end of all known galaxies.  As such they present a number of challenges for any theory of galaxy formation \citep{Weinberg2015}. 
Their typically high dynamical-mass-to-light ratios \citep{Mateo1998} indicate the presence of massive dark matter haloes. The velocity dispersion and half-light radius of dSphs are approximately related as $\sigma_\mathrm{los}\propto R_h^{1/2}$ which can be interpreted as all dSphs residing in a \emph{universal} dark matter halo \citep{Strigari2008,Walker2009,Wolf2010,Sawala2016}
corresponding to the smallest dark matter halo that can form and retain stars in $\Lambda$CDM galaxy formation theory \citep{OkamotoFrenk2009}. From the classical dSphs, this smallest dark matter halo has peak circular velocity $V_\mathrm{max}\sim18\kms$ with an uncertainty / natural scatter of $\sim3\kms$ \citep{Sawala2016}.

However, a number of dSphs (in both the Milky Way and M31) fall significantly under this universal halo relation. 
The most discrepant of these (in the Milky Way) is the recently-discovered `feeble giant' dwarf spheroidal galaxy (dSph) Crater II \citep[$R_h\approx1.1\,\mathrm{kpc}$, $L_V\approx1.5\times10^5\,\mathrm{L}_\odot$,][]{Torrealba2017}. This dSph appears problematic to fit into standard galaxy formation theory due to its extreme kinematic coldness \citep[$\sigma_\mathrm{los}\approx2.6\,\mathrm{km\,s}^{-1}$,][]{Caldwell2017} 
compared to an expected $\sigma_\mathrm{los}\approx11\kms$. The observations of Crater II are entirely in line with the predictions of Modified Newtonian Dynamics \citep[MOND, $\sigma_\mathrm{los}\approx2\,\mathrm{km\,s}^{-1}$,][]{McGaugh2016} in which a low velocity dispersion (anticipated for low luminosity dwarfs) is further suppressed by the external field effect from the Galaxy \citep[expected when the internal acceleration is significantly smaller than the external,][]{Milgrom1983,McGaughMilgrom2013}.

Cosmological simulations of dSphs \citep{Sawala2016,Frings2017,Fattahi2017} demonstrate that tidal stripping \citep[e.g.][]{Hayashi2003,Penarrubia2008,vandenBosch2017} significantly suppresses the velocity dispersion producing objects similar to Crater II. A further consideration is possible geometric effects \citep[e.g][]{SandersEvans2016a,SandersEvans2016b,SandersEvans2017}, as significant flattening along the line of sight (of both light and dark matter) can suppress the line of sight velocity dispersion. Thirdly, simulations show natural variance in the concentration and maximum circular velocity of haloes hosting dSphs. Here, we investigate whether Crater II can be fit into the standard $\Lambda$CDM galaxy formation picture through a combination of these effects.

We begin our investigation by detailing in Section~\ref{Sec::Properties} the properties of dSphs and discussing the approximate effect of shape and tides. In Section~\ref{Sec::Nbody}, we use these properties to inspire controlled two-component $N$-body simulations to thoroughly inspect the relationship between mass loss, velocity dispersion suppression and shape. By calibrating a simple probabilistic model using the $N$-body simulations, we explore a wider range of possibilities for Crater II putting constraints on its orbital history in Section~\ref{Sec::Prob}. In Section~\ref{Sec::CoreCusp}, we investigate the sensitivity of our conclusions to the dark and light mass profiles, and in particular study the effects of a weak dark matter cusp or core. In Section~\ref{Sec::Conclusions}, we present conclusions and thoughts on future observational tests that could be performed in light of our work.

\section{Dwarf spheroidal properties and evolution}\label{Sec::Properties}

\subsection{Universal relation}
dSphs follow a trend between the line-of-sight velocity dispersion $\sigma_\mathrm{los}$ (luminosity-averaged over the whole dSph) and the half-light major-axis length $R_\mathrm{h}$.
\cite{Walker2010} demonstrates that the dSphs of the MW and M31 fall on the low-mass end of the relationship between circular velocity and radius traced by galaxies over a wide mass range.
Converting this relationship into $\sigma_\mathrm{los}$ gives $\log_{10}\sigma_\mathrm{los}=(1.271^{+0.15}_{-0.19})+0.5\log_{10}R_\mathrm{h}$. 

Another suggestion \citep{Walker2009,Wolf2010} is that all the dSphs live within a universal dark matter halo (the lowest mass halo that admits star formation and subsequently retains the stars). We adopt the results from \cite{Sawala2016} comparing subhaloes in the APOSTLE simulations to the classical dSphs and measure a mean peak circular velocity $V_\mathrm{max}=17.6\kms$ with scatter $\Delta_{V_\mathrm{max}}=3.2\kms$ (accounting for the uncertainties and using the dark-matter-only uncertainties for Draco and Fornax for which there are insufficient APOSTLE analogues to reliably estimate the uncertainty in $V_\mathrm{max}$). Converting the circular velocity to velocity dispersion via $V_c=\sqrt{2.5}\sigma_\mathrm{los}$ \citep{Walker2010}, we have the relation
\begin{equation}
\sigma_\mathrm{los}^2=1.85 V_\mathrm{max}^2\Big[\frac{r_{s\mathrm{DM}}}{r}\Big(\ln(1+(r/r_{s\mathrm{DM}}))-\frac{r}{r+r_{s\mathrm{DM}}}\Big)\Big],
\label{eqn::universal}
\end{equation}
which for small $r$ (assuming a concentration $c=20$ halo \footnote{Our convention is $M(<cr_{s\mathrm{DM}})=\Delta\tfrac{4}{3}\pi(cr_{s\mathrm{DM}})^3\rho_\mathrm{crit}$ with $\Delta=101.1$, $\rho_\mathrm{crit}=3H_0^2/(8\pi G)$ 
and $H_0=67.8\kms\,\mathrm{Mpc}^{-1}$} \citep{Maccio2008} with $V_\mathrm{max}=17.6\kms$ and using $r_h=(4/3)R_\mathrm{h}$) 
gives
$\log_{10}\sigma_\mathrm{los}=1.24+0.5\log_{10}R_\mathrm{h}$ -- entirely consistent with the results of \cite{Walker2010}. For dSphs with half-light radii close to the radius of maximum circular velocity (such as Crater II), the predicted velocity dispersion is quite insensitive to choice of concentration. However, we consider the scatter in the concentration taken from \cite{Maccio2008} of $\Delta_{\log_{10}c}=0.13$ for haloes of mass $\sim3\times10^9\mathrm{M}_\odot$.  
We use this latter universal relation and assume that the relationships are followed by spherical non-tidally-stripped dSphs. When applied to Crater II (using $R_\mathrm{h}=1.1\,\mathrm{kpc}$), this formula gives a velocity dispersion suppression of $\sim0.24$ relative to the universal relation.

\subsection{Shape}

We also consider the possibility of scatter about these relationships due to geometric effects. For simplicity, we consider `equally-flattened' dwarf galaxies i.e. those for which the stars and dark matter are stratified on concentric self-similar ellipsoids with axis ratios $p=b/a$ and $q=c/a$. There is some evidence for such a simplification from simulations \citep{Knebe2010}. When viewed at standard spherical polar angles $\theta$ and $\phi$ relative to the Cartesian coordinate system defined by the principal axes, the observed velocity dispersion and on-sky half-light radius can be computed using simple analytic expressions \citep{Roberts1962,SandersEvans2016b}. For the current application, it is worth considering the oblate spheroidal case, as the observed ellipticity of Crater II is consistent with being round ($\epsilon\lesssim0.1$). When viewing an oblate spheroid `face-on' ($\theta=0$), the measured velocity dispersion is smaller than the spherical average. For instance, if the axis ratio is $q=0.3$, the suppression factor is $\sim0.67$. In a more extreme case, if $q=0.1$, the factor is $\sim0.41$. Additionally, the measured half-light radius is larger than the spherically-averaged half-light radius as the galaxy has been stretched on the sky (and compressed along the line-of-sight). For $q=0.3$, the half-light radius is larger by a factor $\sim1.5$. Both of the effects produce scatter away from any universal relation between velocity-dispersion and radius in the same sense as the discrepancy of Crater II. However, only very extreme flattening and a fortuitous viewing down the minor axis can produce the degree of velocity dispersion suppression and shape of Crater II. We require a further mechanism to contribute to the suppression.

\subsection{Tidal disruption}\label{Section::Tidal}

Tidal disruption removes mass from the centre of a dwarf galaxy producing a suppression in the velocity dispersion. The tidal radius of a spherical galaxy with mass profile $M(r)$ with instantaneous orbital frequency $\Omega=\mathrm{d}\phi/\mathrm{d}t$ at radius $r$ in a spherical Galactic potential $\Phi$ is \citep{King1962,vandenBosch2017}
\begin{equation}
r_\mathrm{t}(r)=\Big(\frac{GM(r_\mathrm{t})}{\Omega^2-\partial_r^2\Phi}\Big)^{\frac{1}{3}}.
\label{eqn::tidal}
\end{equation}
For a galaxy on a circular orbit in a gravitational field with mass profile $M_G(r)$ and logarithmic density slope $-\gamma$, this equates to
\begin{equation}
r_\mathrm{t}(r)=\Big(\frac{M(r_\mathrm{t})}{\gamma M_G(r)}\Big)^{\frac{1}{3}}r,
\label{eqn::tidal_radius}
\end{equation}
On more eccentric orbits, the centrifugal term increases relative to the force gradient but in the near radial limit, the force gradient wins out so the denominator becomes $(\gamma-1)$.
In the tidal approximation \citep{vandenBosch2017}, in which all material outside the tidal radius is instantaneously stripped, a galaxy with pericentric radius $r_\mathrm{p}$ and apocentric radius $r_\mathrm{a}$, will have mass ratio after a single pericentric passage equal to
\begin{equation}
\frac{M}{M_0}=\frac{M(r_\mathrm{t}(r_\mathrm{p}))}{M(r_\mathrm{t}(r_\mathrm{a}))}.
\label{eqn::massloss}
\end{equation}
Applying this formula repeatedly for $N$ pericentric passages results in severe overestimates of the mass loss \citep{Hayashi2003} as the dark matter does not simply settle back to its original distribution with depleted mass. Additionally, the tidal radius decreases with increasing mass loss. \cite{Hayashi2003} argue that once NFW profiles become truncated they follow the law
\begin{equation}
\rho(r)=\rho_{\mathrm{NFW}}(r)\frac{f_\mathrm{te}}{1+(r/r_\mathrm{te})^3},
\end{equation}
and give fits for $f_\mathrm{te}$ and $r_\mathrm{te}$ as a function of the ratio of the currently bound mass to the initial mass. An improved estimate of the total mass loss can be obtained using this simple formalism. After each pericentric passage, the tidal radius and hence the mass loss at the next pericentre is computed using equations~\eqref{eqn::tidal} and~\eqref{eqn::tidal_radius} (with both factor $\gamma$ and $\gamma-1$). The solutions must be found iteratively. As shown by \cite{Hayashi2003}, this formalism underestimates the mass loss from the dSph (possibly as there are energetic particles inside the tidal radius that also become unbound) in a way that varies with the orbital properties. The discrepancy is largest (factor of three) for highly eccentric ($e=0.75$) orbits where a superior method for computing the mass loss is the impulse approximation \citep{AguilarWhite1986,vandenBosch2017} although its (repeated) application is more computationally costly \citep{Hayashi2003}. 

Both \cite{Hayashi2003} and \cite{Penarrubia2008} demonstrate that the dSph's structural and dynamical evolution are governed solely by the cumulative mass loss. With the total mass loss within the initial projected half-light radius computed by the above method, the change in (spherical) velocity dispersion and half-light radius are given by the expressions in \cite{Penarrubia2008}. The ratio of each property to its initial value follows a simple relation
\begin{equation}
g(x) = \frac{2^\alpha x^\beta}{(1+x)^\alpha}
\label{eqn::gx_penarrubia}
\end{equation} 
with $x=M_i/M_{i0}$ where $M_i$ is the total mass within a fixed radius $R_i$ (for which \cite{Penarrubia2008} uses the King radius $R_c$). We transform the relations to use the total mass $M_\mathrm{h}$ within the initial projected half-light radius $R_\mathrm{h}(t=0)$ and using the relations for King radius and concentration compute the evolution of the half-light radius \citep{Fattahi2017}. The resulting fit parameters are given in Table~\ref{Table::PMN08AlphaBeta}. When discussing mass loss, we mean the dark plus stellar mass loss although the dark mass is 
always dominant for our purposes.
\begin{table}
\caption{Fitting parameters for dSph properties following equation~\eqref{eqn::gx_penarrubia}. $R_\mathrm{h}^\mathrm{PMN08}$ is computed from the results of \protect\cite{Penarrubia2008} whilst the others are calibrated to our simulations.}
\begin{center}
\begin{tabular}{cccccc}
&$\sigma_\mathrm{los}$&$R_\mathrm{h}^\mathrm{PMN08}$&$R_\mathrm{h}$&$(M/L)$&$L$\\
\hline
$\alpha$&-0.12&2.3&0.6&-2&1\\
$\beta$&0.45&0.6&0.45&-1.2&2.2\\
\hline
\end{tabular}
\end{center}
\label{Table::PMN08AlphaBeta}
\end{table}

To explain the factor $0.24$ velocity dispersion suppression observed in Crater II solely from tidal disruption, we require Crater II to have lost $\sim95\percent$ of its central total mass. This level of mass loss only suppresses the half-light radius by a factor $\sim0.7$ so the net effect is to move Crater II downwards in the $\sigma_\mathrm{los}$ against $R_\mathrm{h}$ plane. Such analytic arguments suggest that, given Crater II started life on the universal dSph relation, its current observables can be explained by heavy tidal mass loss. Non-sphericity aids in suppression of velocity dispersion but cannot account for the entirety of the effect.

\section{Controlled N-body simulations}\label{Sec::Nbody}

We have suggested a possible formation pathways for Crater II. We now turn to constructing a number of N-body simulations for the evolution of Crater II that more accurately reflect the effects of flattening and tidal disruption. 

\subsection{Setup}

We consider a set of Plummer stellar profiles (denoted with $\star$) embedded in NFW dark matter haloes (denoted by DM). Both stars and dark matter follow the truncated double-power-law density law
\begin{equation}
\rho_i(r)\propto r^{-\gamma_i}\Big(1+(r/r_{si})^{\alpha_i}\Big)^{-(\beta_i-\gamma_i)/\alpha_i}\mathrm{sech}(r/r_\mathrm{t}),
\end{equation}
where we choose $(\alpha,\beta,\gamma)_\mathrm{DM}=(1,3,1)$ for an NFW profile and $(\alpha,\beta,\gamma)_\star=(2,5,0)$. Although the surface brightness profiles of several classical dSphs (most notably Fornax) are well fit by \cite{King1962} profiles, the majority are well approximated by Plummer profiles (e.g. Sculptor). For the ultrafaint dSphs, the data is too poor to distinguish between these profiles. Additionally, \cite{Penarrubia2009} has demonstrated that, after some tidal stripping, the surface profiles of simulations tend to appear more Plummer-like. However, it should be noted that a weakly-cusped profile projects to a cored profile. The dark matter profile is specified by the concentration $c=20$ \citep{Maccio2008}
and the choice of scale radius $r_{s\mathrm{DM}}$. We choose to work with a universal dark-matter profile for dSphs with a maximum circular velocity of $20\kms$ \citep{Walker2009,Sawala2016} resulting in a dark-matter scale-radius of $r_{s\mathrm{DM}}=1.45\kpc$. To specify the stellar profile, we define a segregation parameter \citep{Penarrubia2008} as
\begin{equation}
s\equiv r_{s\mathrm{DM}}/r_{s\star}.
\end{equation}
We adopt a simple relation between the stellar mass and mass-to-light ratio \citep{Mateo1998,Fattahi2017} of $M_\mathrm{DM}(<r_{h,\star})/(M_\star/2) = 200(10^5\mathrm{M}_\odot/M_\star)^{0.4}$ (assuming a stellar-mass-to-light ratio of $1$).
The tidal radius $r_\mathrm{t}$ is computed using equation~\eqref{eqn::tidal_radius} at the initial orbital radius (assuming the dark matter in the dwarf galaxy dominates and choosing $\gamma=2$). 

With these choices, our density models are completely specified up to the choice of $s$. We also consider oblate models specified by the flattening parameter $q$ where $r^2\equiv q^{2/3}(R^2+(z/q)^2)$. 
To construct dynamical equilibria corresponding to these distributions, we specify the constant anisotropies as $\beta_\mathrm{DM}=0$ and $\beta_\star=-0.2$ (a weak tangential bias is necessary to produce a cored Plummer profile in a cusped central potential, see \cite{AnEvans2006}). The corresponding distribution functions (assuming sphericity) are computed numerically \citep{Cuddeford1991} and $N_\mathrm{DM}$ dark matter particles and $N_\star$ star particles are sampled. Flattening is introduced by rescaling the spatial coordinates of the model and adjusting the velocities to satisfy the tensor-virial theorem \citep{Dehnen2009}. All models are then run for $35-45$ orbital time units in a Made-to-Measure algorithm \citep{Dehnen2009} using the coefficients of a potential-density basis function expansion \citep{Zhao1996}. Note we do not fix the kinematics, so the anisotropy is allowed to vary away from the initial guess model. Dark matter and stars are run separately in their combined potential decomposed using the basis function expansion. The results are stacked, scaled to physical units and shifted to the initial orbital phase-space location. We do not rotate the flattened model such that it is initially aligned with the Milky Way disc.

\subsubsection{Orbital properties}\label{Section::Orbit}

We place our model dwarf galaxies on plausible Crater II orbits.
From \cite{Torrealba2017}, the distance is $117.5\pm1.1\,\mathrm{kpc}$ whilst \cite{Caldwell2017} find the line-of-sight velocity as $87.5\pm0.4\,\mathrm{km\,s}^{-1}$. There is considerable uncertainty on the proper motion of Crater II. \cite{Caldwell2017} model the variation of the line-of-sight velocities across Crater II to obtain an estimate of the proper motion as $(\mu_{\alpha*},\mu_\delta)=(-0.18\pm0.16,-0.14\pm0.19)\mathrm{mas\,yr}^{-1}$ (although this variation is entirely degenerate with any internal dSph rotation, if present). We consider the possible orbits of Crater II in the Milky Way potential from \cite{McMillan2017}. We ignore the effects of dynamical friction which is anticipated to be negligible for a halo with virial mass $1\times10^9\,\mathrm{M}_\odot$. Due to its measured line-of-sight velocity it is impossible for Crater II to be on low eccentricity orbits with $e\lesssim0.4$. Although our potential contains an axisymmetric disc component, Crater II primarily orbits in regions of the Galaxy where the (model) potential is spherical and so its orbital motion is nearly completely determined by the magnitude of the transverse velocity. We parametrize this using the magnitude of the proper motion $|\mu|$\footnote{In a slight abuse of notation, $|\mu|$ denotes the magnitude of the on-sky proper motion \emph{vector}.} relative to the Galactic standard of rest i.e. $|\mu|=0$ produces a near radial orbit and the corresponding proper motion of Crater II would be $(\mu_{\alpha*},\mu_\delta)=(-0.206,-0.241)\mathrm{mas\,yr}^{-1}$ (due entirely to the solar reflex motion). We rewind the orbit for $13.7\,\mathrm{Gyr}$ and place the equilibrium galaxy at the last apocentre encountered. 

\subsubsection{Simulation}

The dwarf galaxy is evolved forwards using \texttt{griffin} -- a parallelized $N$-body code \citep{Dehnen2000,Dehnen2002} using a fast multipole method based on \cite{Dehnen2014}. We adopt a softening length according to the prescription in \cite{Dehnen2001} for a Hernquist sphere of $N$ particles modified by the factors given in the \texttt{gyrfalcON} manual for a P1 kernel and scaled to $r_{-2}$ for an NFW model: $\epsilon=0.048r_{s\mathrm{DM}}(N/10^5)^{-0.23}$. We use individual time-steps ensuring that the time-step is not greater than $0.2\sqrt{\epsilon/a_i}$ or $0.2 \sqrt{\Phi_i}/a_i$. For small numbers of particles, the central potential produced by the dark matter and experienced by the stars has considerable numerical noise causing the stellar profile to broaden, settling to $s=1$. 
Choosing $N_\mathrm{DM}=N_\star=4\times10^5$ produces a $s=5$ dwarf galaxy that does not evolve significantly over time (simulations of more segregated dwarf galaxies (e.g. $s=10$) with higher numbers of particles would be necessary for understanding e.g. Draco II, Segue 1 or Grus II). This choice produces energy conservation to $0.1\percent$ over a Hubble time for an isolated $s=2$ galaxy and total energy conservation to $\lesssim0.5\percent$ relative to the initial internal energy of the dSph (defined by $W/2$ where $W$ is the internal potential energy). Additionally, we have run a test with $N_\mathrm{DM}=N_\star=4\times10^6$ and $s=2$ on the most eccentric orbit we consider and find the evolution of half-light radius, central mass and velocity dispersion are very similar to the simulations with fewer particles except when the dwarf galaxy is nearly completely destroyed when the central mass loss is overestimated in the simulation with fewer particles. Additionally, the central shapes of the dwarf galaxies can be overestimated by axis ratio differences of at most $0.1$ in the smaller simulation.

Our models are defined by three parameters: the segregation $s$, the flattening $q$ and the magnitude of the proper motion $|\mu|$ relative to the proper motion that produces no transverse velocity at the current time (i.e. that produced by the solar reflex). We consider a small grid of these parameters given by
\begin{enumerate}
\item $s = 0.5, 1, 2$,
\item $q = 1, 0.3$,
\item $|\mu| = 0.05, 0.1, 0.17, 0.25 \masyr$.
\end{enumerate}

In Table~\ref{Table::OrbitalProperties}, we give the properties of the four orbits considered.
\begin{table}
\caption{Orbital properties of our simulations. $r_{\mathrm{p}},r_{\mathrm{a}}$ are the pericentre and apocentre, $T_r$ the radial period and $r_\mathrm{t}(r_\mathrm{p})$ the tidal radius at pericentre.}
\begin{tabular}{ccccc}\hline$|\mu|/\,\mathrm{mas\,yr}^{-1}$&$r_\mathrm{p}/\,\mathrm{kpc}$&$r_\mathrm{a}/\,\mathrm{kpc}$&$T_r/\,\mathrm{Gyr}$&$r_\mathrm{t}(r_\mathrm{p})/\,\mathrm{kpc}$\\\hline0.05&4.99&130&1.64&0.139\\0.1&13&131&1.75&0.636\\0.17&28.4&135&1.99&1.79\\0.25&52&146&2.48&3.53\\\hline\end{tabular}
\label{Table::OrbitalProperties}
\end{table}

\subsection{Analysis}

\begin{figure}
$$\includegraphics[width=\columnwidth]{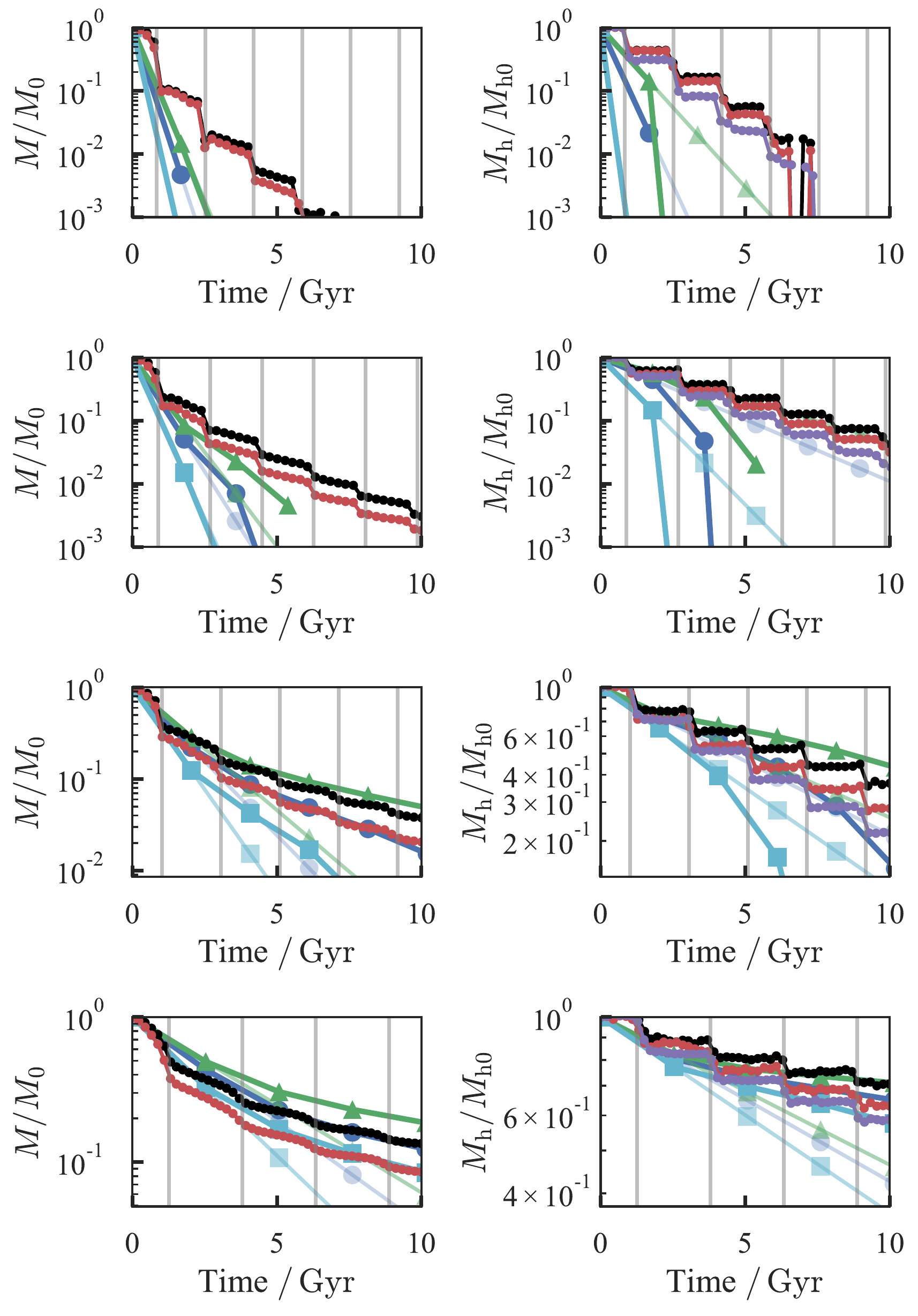}$$
\caption{Tidal mass loss: the left set of panels show the total bound dark matter mass normalized to the initial mass (black spherical, red flattened) whilst the right shows the mass within the initial half-mass radius $R_\mathrm{h}=\tfrac{1}{2}r_{s\mathrm{DM}}$ (and inside $r_{s\mathrm{DM}}$ for the spherical model in purple). Each row corresponds to a different proper motion of Crater II today in the Galactocentric rest frame. The solid lines with large symbols show the predicted mass loss from the tidal approximation formalism of \protect\cite{Hayashi2003} (inside $\tfrac{1}{2}r_{s\mathrm{DM}}$ for the right panels). Green triangles show the results of equation~\eqref{eqn::tidal_radius}, blue circles using $(1-\gamma)$ in the denominator and cyan square using equation~\eqref{eqn::tidal}. Faint lines show repeated application of initial mass loss. Grey vertical lines mark pericentric passages.}
\label{Fig::MassLoss}
\end{figure}

At each stored time-step (approximately every $300\,\mathrm{Myr}$), we find the centre of the remnant dwarf spheroidal using the \texttt{bound\_centre} method described in the \texttt{gyrfalcON} manual. This method seeks the most bound particle (ignoring the velocities) along with its $K=512$ nearest neighbours and computes the phase-space centre by averaging the $K/8$ most bound weighted by their energy. We remove all unbound particles and recompute the potential of all particles and the new centre, iterating until number of bound particles changes by less than $20$. For the stars and dark matter separately, we diagonalize the reduced moment of inertia tensor (computed iteratively using an ellipsoidal radius) to find the axis ratios and principal axes of the entire distribution. We find the velocity dispersions of the stars along each of the principal axes. We also compute the density associated with each particle using a kernel density estimate and split the particles ordered by density into at most $30$ bins of at least $30$ particles, for each of which we compute axis ratios and the principal axes from the moment of inertia. We compute a Plummer fit to the on-sky distribution of the stellar remnant (finding the half-light major axis length $R_\mathrm{h}$ and the observed ellipticity $\epsilon$) and its projected velocity dispersion $\sigma_\mathrm{los}$ (using velocities projected onto the vector from the observer to the centre of the galaxy to remove the perspective rotation produced by the bulk motion of the galaxy). Finally, we find the bound mass fractions (both total $M/M_0$ and within the original half-light radius $M_\mathrm{h}/M_\mathrm{h0}$).

\subsection{Dynamical evolution}

\begin{figure}
$$\includegraphics[width=\columnwidth]{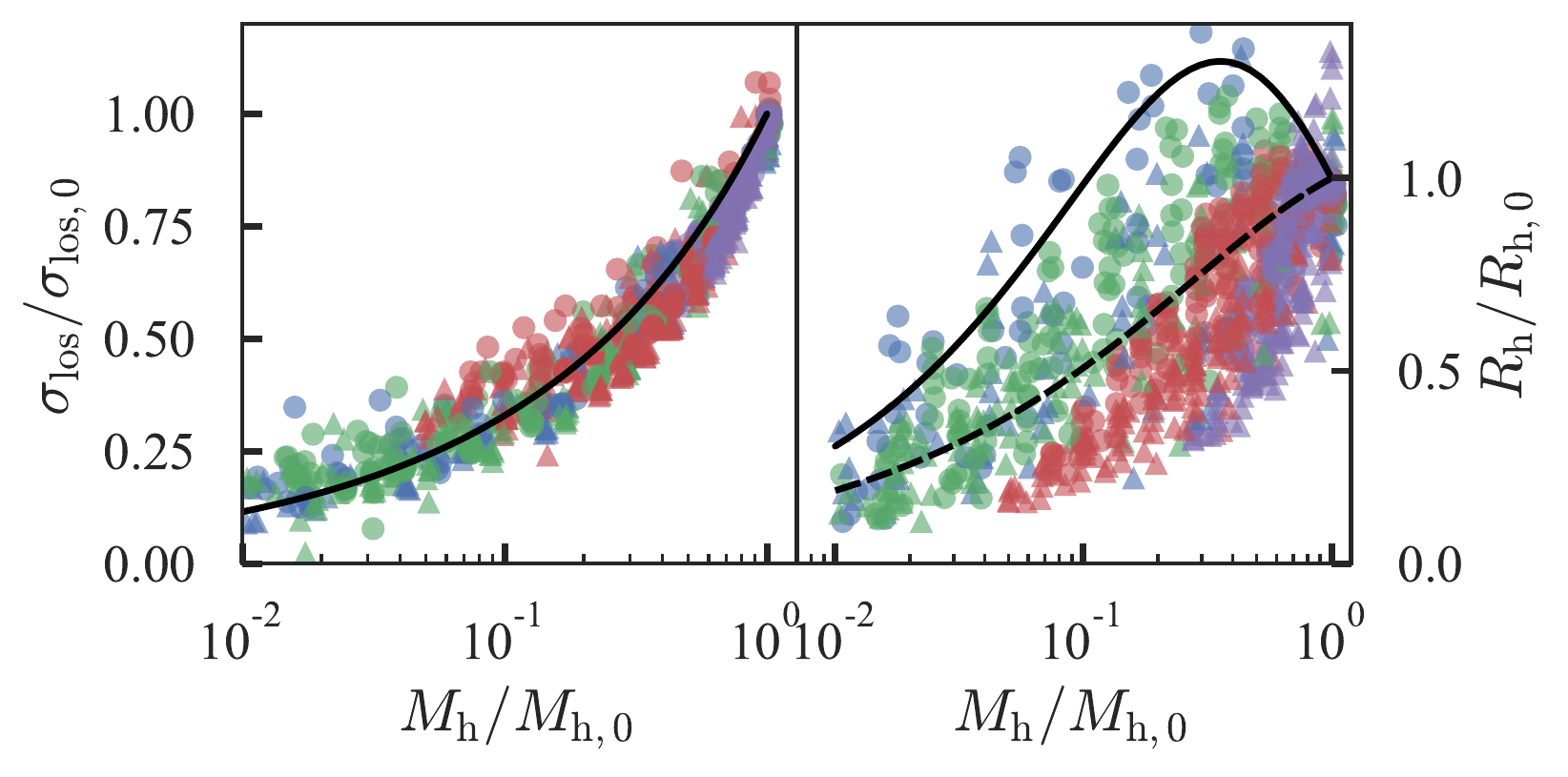}$$
$$\includegraphics[width=\columnwidth]{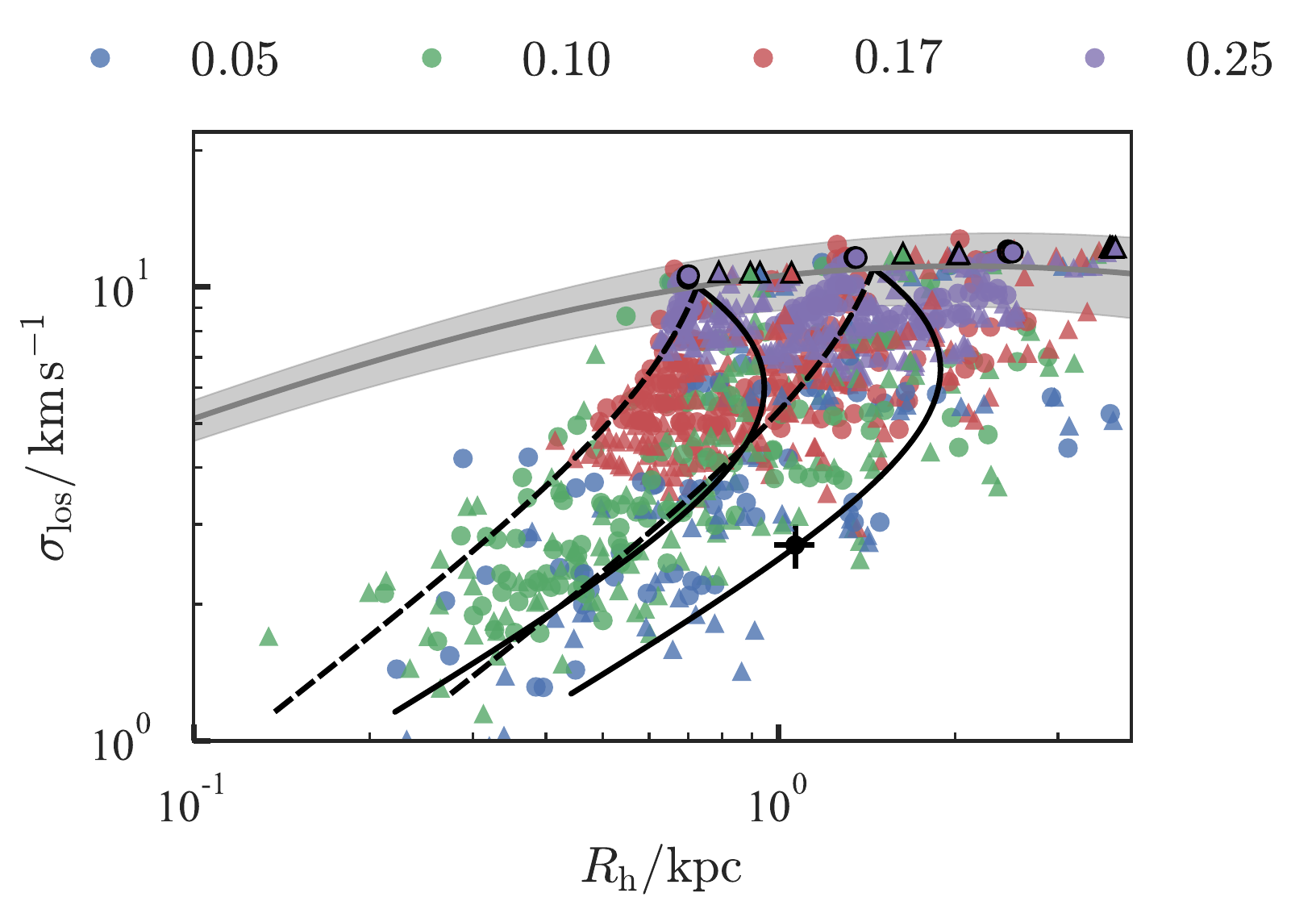}$$
\caption{Evolution of projected velocity dispersion and half-light radius with central mass remnant. Each panel displays the results from each time-step of the simulations (initial time-step outlined) with $s=0.5, 1, 2$ coloured by $|\mu|$ (triangles are initially flattened models and circles spherical). The black lines show the tracks from \protect\cite{Penarrubia2008}, the dashed black from our approximate tracks and the grey band the expected relationship if all dSphs live within similar dark matter haloes ($V_\mathrm{max}=(17.6\pm3)\,\mathrm{km\,s}^{-1}$ and $c=20$). The black point shows the observations of Crater II.}
\label{Fig::SlosRh}
\end{figure}

In Figure~\ref{Fig::MassLoss}, we show the fraction of remaining mass in the simulations (both total and inside the initial half-light radius). Both total and central mass are discontinuous around pericentric passage (particularly for small $|\mu|$) reflecting the impulsive nature of tidal stripping.
However, between pericentric passages the total mass smoothly declines, whilst the mass within the half-light radius is approximately constant, demonstrating that the central material remains `protected' between pericentres.
We also observe that there is increased mass loss for the flattened set of simulations although the difference is smaller for the simulations with more radial orbits. The tidal mass loss using the analytic formalism presented in Section~\ref{Section::Tidal} is shown with the blue line. For more tangential orbits ($|\mu|=0.25, e\approx0.5$), the analytic prediction accurately gives both the total and central mass loss (although as found by \cite{Hayashi2003} equation~\eqref{eqn::tidal_radius} slightly underestimates the total mass loss). For more radial orbits, the analytic prediction grossly overestimates the degree of tidal stripping. From the $s=2$ simulations, we compute the mass fraction remaining per pericentric passage as a function of $|\mu|$ which is well approximated by\footnote{Another fitting function which could be used for a wider array of possible orbits is $M_\mathrm{h}/M_\mathrm{h0}=1/(1+(r_\mathrm{a}/r_\mathrm{p})/16)^{1.2}$ (parameters are $11$ and $1.2$ for the $s=1$ simulations) but we have not tested this on orbits not consistent with Crater II.}
\begin{equation}
M_\mathrm{h}/M_\mathrm{h0}=1-\exp(-|\mu|/0.11\,\mathrm{mas\,yr}^{-1}).
\label{eqn::Mmufit}
\end{equation}
Using the $s=1$ simulations changes the scale to $0.14\,\mathrm{mas\,yr}^{-1}$ and similarly $0.17\,\mathrm{mas\,yr}^{-1}$ for the $s=0.5$ simulations. We further ran a simulation with $|\mu|=0.42\,\mathrm{mas\,yr}^{-1}$ and $s=2$ to check the validity of our approximation for orbits that put Crater II near pericentre today. Our approximation gives $2\percent$ mass loss per pericentric passage whilst we find $4\percent$ from the simulations, so our approximation will not bias our later results.

In Figure~\ref{Fig::SlosRh}, we show the corresponding evolution of the velocity dispersion and observed half-light radius. As \cite{Penarrubia2008} found, the velocity dispersion follows a universal trend with central mass irrespective of orbital history and segregation. Additionally, for our study we have demonstrated that this track is also followed by flattened dwarfs and weakly embedded systems ($s=0.5, 1$ -- \cite{Penarrubia2008} considered $s=2,5,10$). However, the corresponding evolution of the half-light radius is not a simple function of central mass. \cite{Penarrubia2008} used King models for the stars and found that both the King radius and concentration increased, resulting in an initial increase in half-light radius before a decline. We use Plummer models for the stars and find this initial increase approximately follows the
track from \cite{Penarrubia2008} only for our most radial orbit, whilst for more circular orbits the half-light radius declines. Figure~\ref{Fig::SlosRh} demonstrates that the half-light radius is not simply a function of the central mass loss changing between pericentric passages. The velocity dispersion rapidly decreases at each pericentric passage as it is most sensitive to the properties of the tightly-bound particles. The half-light radius (as determined from Plummer fits) however is more sensitive to the outskirts of the galaxy. As discussed by \cite{AguilarWhite1986} and \cite{Penarrubia2009}, a tidal impulse produces a set of new weakly-bound particles ejected from the central regions which dynamically mix on longer timescales. This results in a step in the outer profile which steadily moves outwards altering the shape of the best-fitting density profile over an orbital period. We have confirmed that we obtain similar results for the half-light radius using \cite{King1962} profile fits.

Turning to the $\sigma_\mathrm{los}$ against $R_\mathrm{h}$ plane, we observe that our set of simulations produce only a few objects consistent with Crater II (black cross in Fig.~\ref{Fig::SlosRh}).
For $s=2$ our simulations broadly agree with the tidal tracks of \cite{Penarrubia2008}. However, for $s=0.5 $ and $1$, the tidal tracks are too steep -- suppression of velocity dispersion is accompanied by suppression of the half-light radius for weakly-embedded systems. We opt to fit an alternative track to the relationship between $R_\mathrm{h}$ and $M_\mathrm{h}/M_\mathrm{h0}$ for the $s=1$ simulations. This relation is shown by the dashed line in Fig.~\ref{Fig::SlosRh} and the parameters are given in Table~\ref{Table::PMN08AlphaBeta}. Our relation shows that for weakly embedded dSphs in NFW-like halos, low velocity dispersions are necessarily accompanied by low half-light radii. We explore the dependence of this conclusion on more general dark-matter profiles in Section~\ref{Sec::CoreCusp}.

\subsection{Shape evolution}

\begin{figure}
$$\includegraphics[width=\columnwidth]{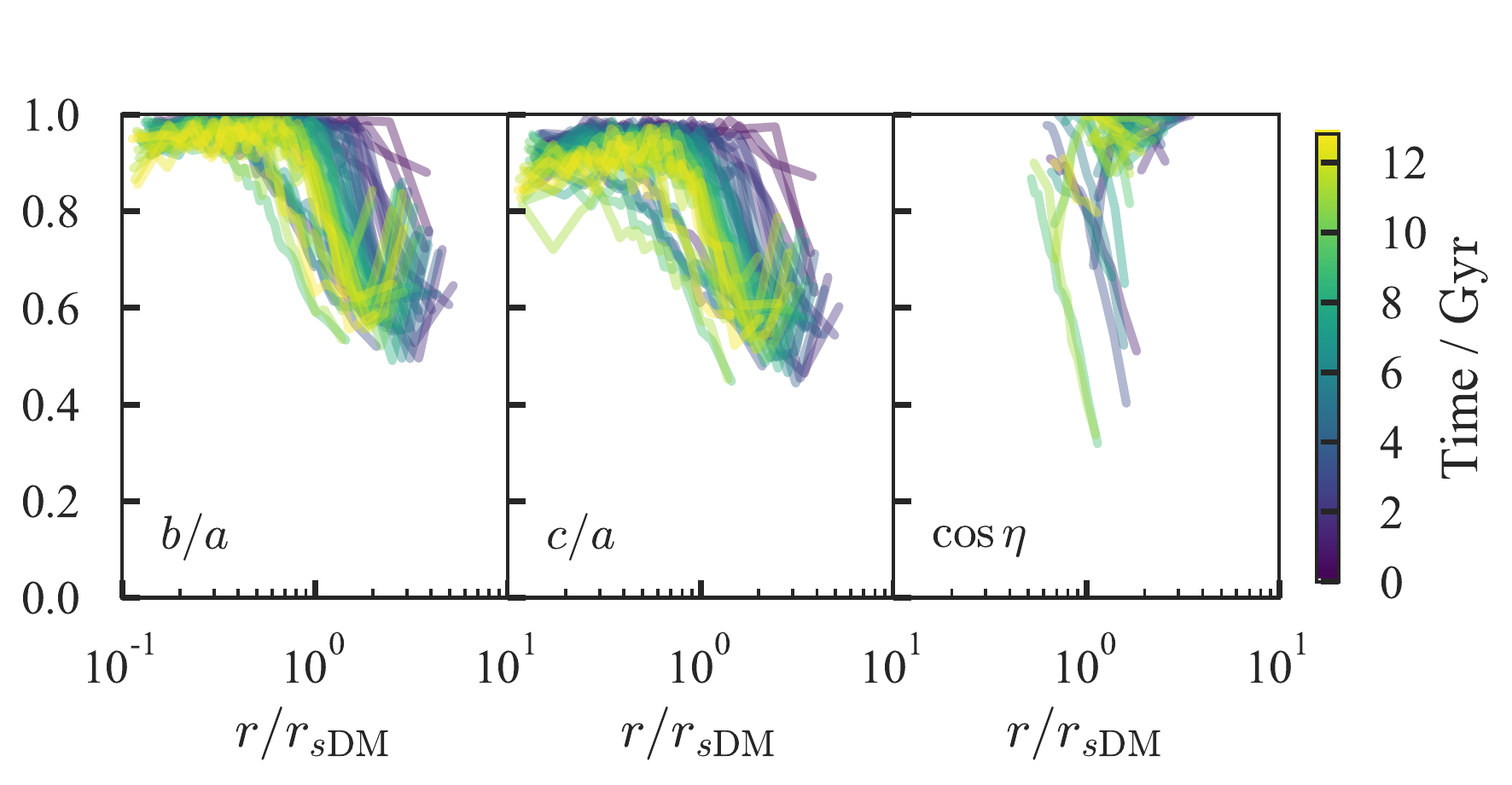}$$
\vspace{-0.8cm}
$$\includegraphics[width=\columnwidth]{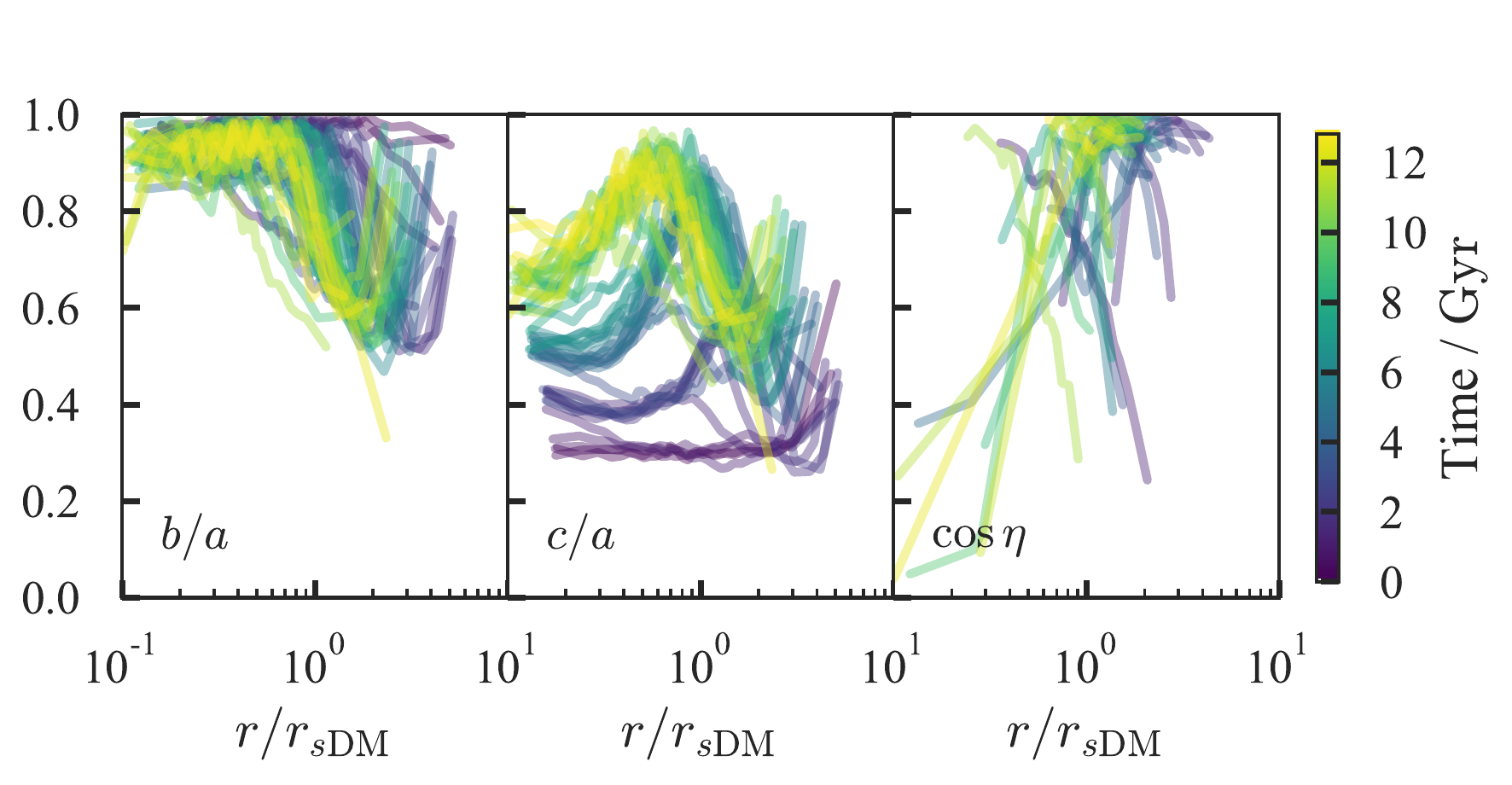}$$
\vspace{-0.8cm}
$$\includegraphics[width=\columnwidth]{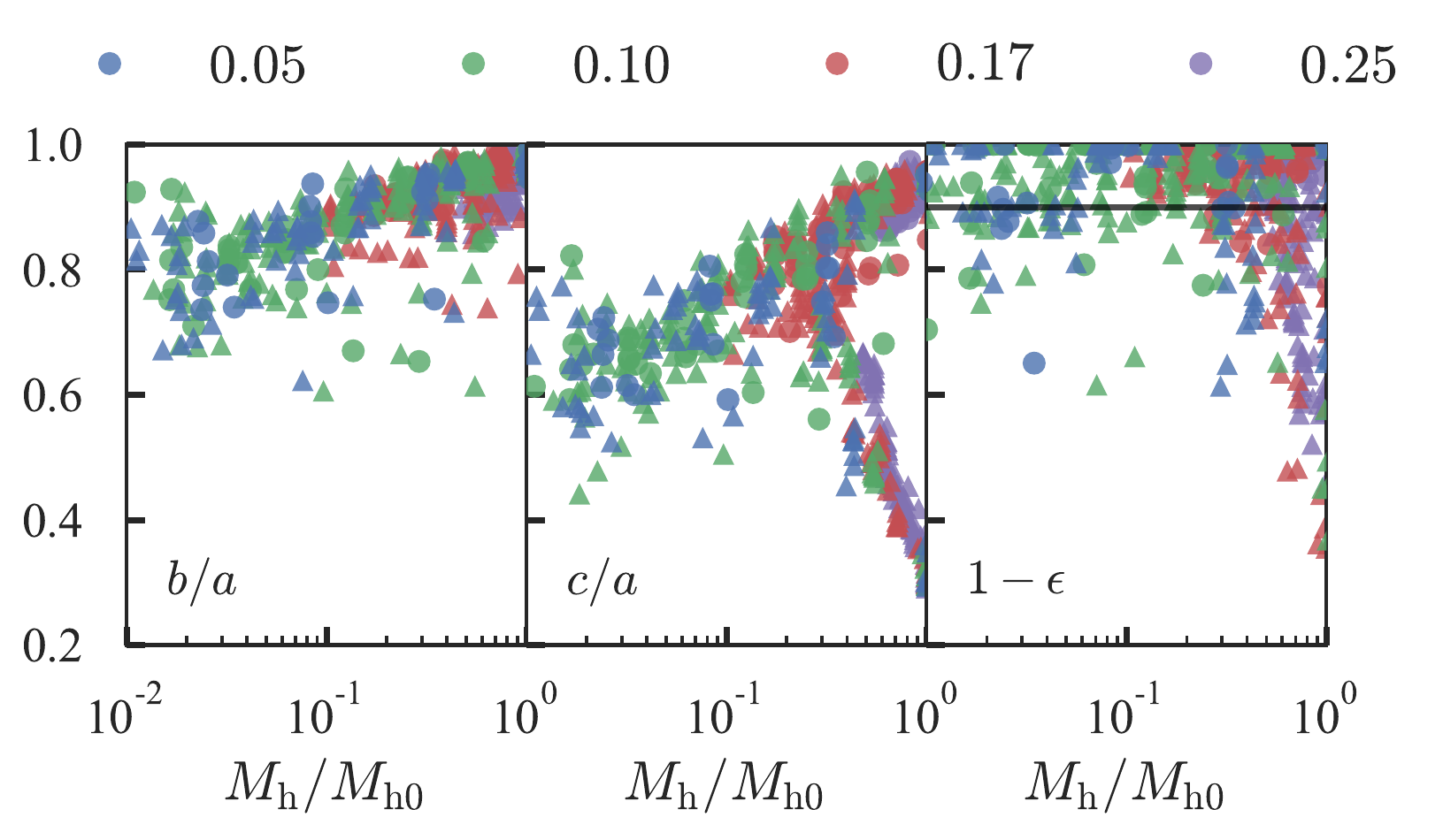}$$
\caption{Tidal shape evolution: \textbf{top and middle}: axis ratios and alignment of major axis with Milky Way centre (only shown where $b/a<0.85$) coloured by time for $|\mu|=0.17\,\mathrm{mas\,yr}^{-1}$. The top row of panels is for the spherical model, bottom the flattened model. \textbf{Bottom}: central ($r<\tfrac{1}{2}R_\mathrm{h}$) axis ratios and on-sky axis ratio as a function of remnant mass for $s=1,2$ coloured by $|\mu|$ with initial spherical models as circles and initial flattened models as triangles (the $90\percent$ confidence limit of Crater II $1-\epsilon$ is shown with a black line).}
\label{Figure::ShapeEvolution}
\end{figure}

In Figure~\ref{Figure::ShapeEvolution}, we show the detailed shape profiles for the $s=2$, $|\mu|=0.17\,\mathrm{mas\,yr}^{-1}$ simulations with initial $c/a=1$ and $c/a=0.3$. Additionally, we show the alignment between the major axis and the Galactocentric radial vector $\cos\eta=\hat{\boldsymbol{x}}\cdot\hat{\boldsymbol{r}}$ (where $b/a<0.85$). We notice that for the spherical model the outskirts very rapidly become prolate $b/a=c/a=0.6$ and radially aligned with $\cos\eta\approx1$ \citep[c.f.][]{Barber2015}. The interior (within $\sim r_{s\mathrm{DM}}$) remains approximately spherical ($c/a\sim0.9$) and, as such, the major axis and hence radial alignment is ill-defined. The flattening of the prolate density contours varies approximately monotonically from the edge of the spherical interior to the outermost radius. 

For the model that starts oblate ($c/a=0.3$), again we find the outskirts become rapidly prolate (or prolate-triaxial) and radially aligned. The interior remains oblate until pericentric passage at which point the tides near instantaneously reshape the dSph increasing the axis ratio to $c/a=0.4$ at the centre but with a rounder envelope. At the next pericentric passage, the rounder envelope moves further inwards and the central axis ratio increases to $c/a=0.5$. Although this orbit experiences six pericentric passages (see Figure~\ref{Fig::MassLoss}), there appear to only be three (possibly four) discrete steps in central flattening. At the end of the simulation, the outskirts (beyond $r_\mathrm{sDM}$) resemble the spherical model whilst the interior is oblate but much rounder than at the initial time. The effects of tidal shocking has been studied in the context of globular cluster evolution \citep[e.g.][]{GnedinOstriker1997} where `tidal relaxation' ($\langle\Delta E^2\rangle$) operates alongside two-body relaxation to accelerate the core collapse. It appears that tidal shocks on our flattened dwarfs have a similar effect in making the galaxy more isotropic and hence more spherical.

In the lower panel of Figure~\ref{Figure::ShapeEvolution}, we display the mean axis ratios within $\tfrac{1}{2}r_\mathrm{sDM}$ and the observed on-sky axis ratio for all $s=1$ and $s=2$ simulations. Recovering the results of \cite{Penarrubia2008}, we find the axis ratios $b/a$ and $c/a$ depend solely on the mass remnant within $R_\mathrm{h0}$. All models tend towards approximately prolate figures such that at $M_\mathrm{h}/M_\mathrm{h0}=0.01$, $b/a\sim0.8$ and $c/a\sim0.6$. The spherical models steadily become more prolate, whilst the initially oblate models become rounder before connecting to the spherical sequence around $M_\mathrm{h}/M_\mathrm{h0}=0.2$. For the flattened models $1-\epsilon$ initially varies around the orbit with only fortuitous alignments of the minor axis producing observed round figures. However, for $M_\mathrm{h}/M_\mathrm{h0}<0.3$, essentially all observed axis ratios are consistent with Crater II ($\epsilon<0.1$).

From our simulations, we have a consistent picture of the evolution of a flattened satellite galaxy. There are two competing effects driving the shape evolution: tidal shocking and tidal distortion. Tidal shocking is most important for the protected central regions of the cluster and acts to isotropise the velocity distribution and make the galaxy rounder. Tidal distortion reshapes the outskirts of the galaxy where the prolate potential produces a prolate envelope that tidally locks. For extreme mass loss the galaxy core becomes more vulnerable to tidal distortion and irrespective of the tidal shocking the galaxy steadily becomes more prolate. We anticipate that all galaxies irrespective of their initial shape are driven onto this prolate tidal distortion sequence, given sufficient mass loss.

\section{Possible formation channels for Crater II}\label{Sec::Prob}

\begin{figure*}
$$\includegraphics[width=\textwidth]{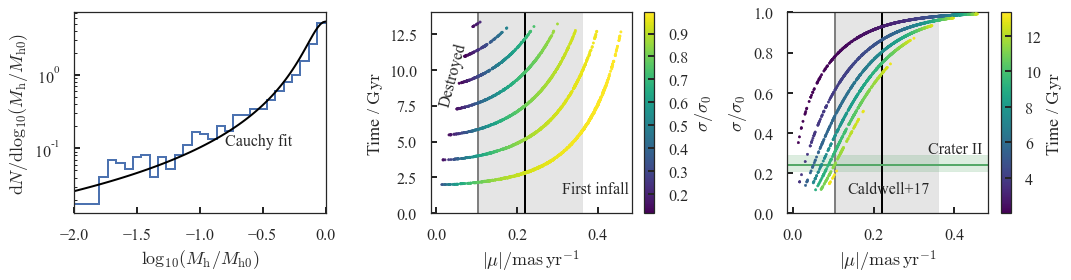}$$
\caption{Possible orbital histories of Crater II: in the left panel we show the distribution of total mass loss within the stellar half-light radius along with an approximate fit using a Cauchy distribution. The middle panel shows the magnitude of the proper motion vector (in the Galactic standard of rest) against total time coloured by final to initial velocity dispersion. The final panel shows a different projection. The black line with grey bands show the proper motion measurement from \protect\cite{Caldwell2017} (with the point estimate given by the grey line). The horizontal green band shows the estimate of Crater II's velocity dispersion suppression.}
\label{Fig::Orbit}
\end{figure*}

We now combine the analytic results on flattening, tidal disruption and the universal dSph relation with the results of our simulations to estimate the most likely configuration of Crater II. Our model consists of a equally-flattened triaxial dSph described by the shape parameters $p=b/a$ and $q=c/a$, along with the two viewing angles $\theta$ and $\phi$. The parent spherical pre-tidally-stripped dSph lies within the scatter of the `universal' relation with a dark matter concentration $c$ and spherically-equivalent half-light radii $R_\mathrm{h0}$ and velocity dispersion $\sigma_\mathrm{los,0}$. Flattening and projection effects adjust these to $R^g_\mathrm{h0}$ and velocity dispersion $\sigma^g_\mathrm{los,0}$. Tidal stripping is parametrized by the mass fraction $M_\mathrm{h}/M_\mathrm{h0}$ remaining within the original half-light radius.

Given this set of parameters, our likelihood is given by
\begin{equation}
\ln\mathcal{L}=-\frac{(\sigma_\mathrm{los}-\sigma'_\mathrm{los})^2}{2\Delta_{\sigma_\mathrm{los}}^2}-\frac{(R_\mathrm{h}-R'_\mathrm{h})^2}{2\Delta_{R_\mathrm{h}}^2}-\frac{\epsilon'^2}{2\Delta_{\epsilon}^2}-\frac{(V_\mathrm{max}-V'_\mathrm{max})^2}{2\Delta_{V_\mathrm{max}}^2},
\label{eqn::likelihood}
\end{equation}
where primed quantities are computed from the model parameters and unprimed denote observations with the $\Delta_i$ quantities giving the associate uncertainties ($V_\mathrm{max}$ `observation' is from \cite{Sawala2016}). We adopt uniform priors on $\ln R^g_\mathrm{h0}$, $\ln\sigma^g_\mathrm{los,0}$, $\cos\theta$, $\phi$ and $p$, and use the prior on $q$ from the analysis of \cite{SandersEvans2017} of $\mathcal{N}(0.45, 0.1)$ (using the notation $\mathcal{N}(\mu,\sigma)$ for a normal distribution with mean $\mu$ and standard deviation $\sigma$). For $\log_{10}c$ we use the prior $\mathcal{N}(1.3,0.13)$ \citep{Maccio2008}.

Given a set of parameters, we compute the observables by first computing the geometric factors (ellipticity $\epsilon'$, the ratio of observed to total velocity dispersion and the ratio of half-light projected major axis length to spherically-equivalent half-light radius) using equations (10-12) of \cite{SandersEvans2016b}. These relate $R^g_\mathrm{h0}$ and $\sigma^g_\mathrm{los,0}$ to $R_\mathrm{h0}$ and $\sigma_\mathrm{los,0}$. From equation~\eqref{eqn::universal} we use $R_\mathrm{h0}$, $\sigma_\mathrm{los,0}$ and $c$ to find $V'_\mathrm{max}$. Given $M_\mathrm{h}/M_\mathrm{h0}$, we use the tidal tracks to relate $R^g_\mathrm{h0}$ and $\sigma^g_\mathrm{los,0}$ to $R'_\mathrm{h}$ and $\sigma'_\mathrm{los}$.

We use the possible orbital histories of Crater II to construct the prior for $M_\mathrm{h}/M_\mathrm{h0}$. The major unknown is Crater II's poorly constrained proper motion. Using the compilation of dSph proper motions from \cite{PawlowskiKroupa2013} complemented with recent proper motions from \cite{CasettiDinescu2017} for Sextans and \cite{Sohn2017} for Draco and Sculptor, we find the total transverse velocity of the dSphs in the Galactocentric rest frame has mean $175\,\mathrm{km\,s}^{-1}$ resulting in a dispersion in each transverse component (assuming isotropy) of $V=140\,\mathrm{km\,s}^{-1}$.
We generate samples in the two transverse velocity components from an uncorrelated Gaussian with variance $V^2$ and integrate the orbits backwards for times $t$ (sampled from a uniform distribution from $0$ to $13.7\,\mathrm{Gyr}$ and choosing the time of the earliest apocentric passage) in the Milky Way potential from \cite{McMillan2017}. We reject samples that imply Crater II is on first infall ($|\mu|\gtrsim0.46\,\mathrm{mas\,yr}^{-1}$), as we have shown tidal mass loss is essential to explain Crater II if it is embedded in an NFW-like dark halo. Figure~\ref{Fig::Orbit} shows the final mass fraction (computed using equation~\eqref{eqn::Mmufit}) and resulting depression in the velocity dispersion (computed using equation~\eqref{eqn::gx_penarrubia}) for our orbit samples. The distribution of the logarithm of the final mass fraction approximately follows a Cauchy distribution with width parameter $0.14$ (using the $s=1$ and $s=0.5$ scales in equation~\eqref{eqn::Mmufit} changes this to $0.22$ and $0.35$ respectively). We use this as a prior in our analysis. Our figure also shows that the \cite{Caldwell2017} measurement of the proper motion implies Crater II is \emph{not} on first infall and to explain Crater II's velocity dispersion suppression \emph{purely by tidal effects} we require $|\mu|\lesssim0.1\masyr$. 

\begin{figure}
$$\includegraphics[width=\columnwidth]{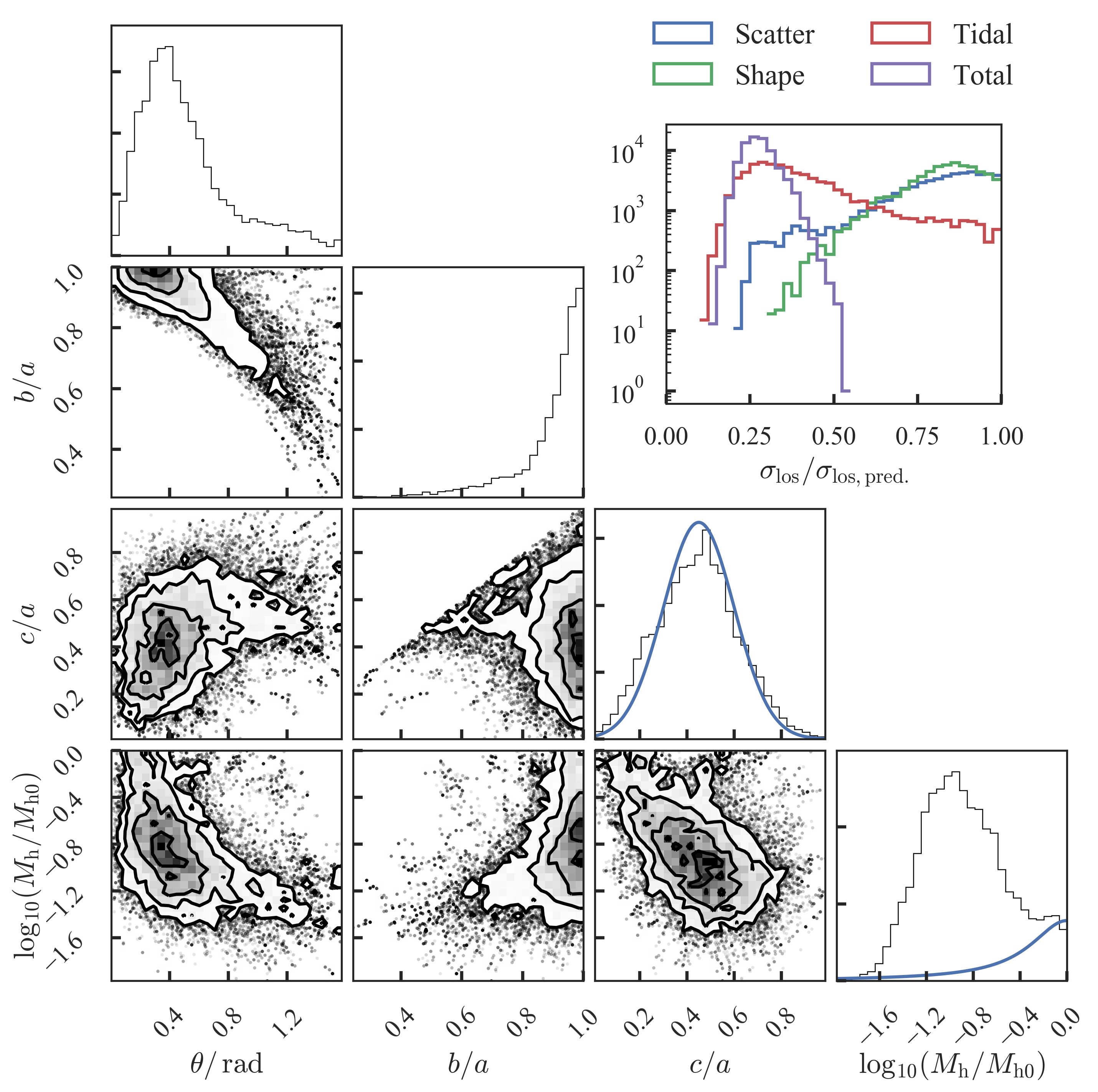}$$
\caption{Posterior samples for four model parameters: the shape parameters $b/a$ and $c/a$, one of the viewing angles $\theta$ (angle relative to the minor axis) and the mass loss fraction. The blue lines show the adopted priors. \textbf{Inset}: Contributions of the three effects to the total suppression of the velocity dispersion of Crater II. In purple we show the total ratio of the true velocity dispersion to its expected dispersion. The three other histograms show the individual effects of natural scatter about the universal relation due to different dark-matter halo properties, the shape and the tidal disruption.}
\label{Figure:Samples}
\end{figure}

In Figure~\ref{Figure:Samples}, we show samples from the posterior defined by the likelihood in equation~\eqref{eqn::likelihood} found using \emph{emcee} \citep{GoodmanWeare2010,ForemanMackey2013}. We observe that $c/a$ shape parameter approximately traces the prior. Highly flattened oblate models viewed near face-on can account for the data with minimal mass loss. Rounder oblate models ($c/a\sim0.4$) can only account for the data if Crater II has lost $90\percent$ of its central dark matter. Prolate models ($b/a\sim c/a$) viewed down the major axis are permitted. These models elevate the velocity dispersion, which must be compensated by increased scatter and tidal disruption. The total mass loss posterior distribution differs significantly from the prior, indicating the properties of Crater II does put constraints on its orbital history. The maximum required mass loss is $\sim97\percent$ of the total.

In the inset of Figure~\ref{Figure:Samples}, we show the contributions of (i) scatter below the universal relation, (ii) shape and (iii) tidal disruption to the overall discrepancy of Crater II's velocity dispersion. We see that tidal disruption produces (on average) the largest suppression of the velocity dispersion (approximately a factor $0.3$) accounting for most of the observed discrepancy. Both flattening and scatter about the universal relation are also necessary ingredients for some scenarios, but contribute approximately a factor $0.8$ each.

Our modelling approach uses a number of simplifications that merit validation: (i) no correlations between flattening and tidal disruption -- we have found that tidal disruption is intrinsically linked to a galaxy's shape (Fig.~\ref{Figure::ShapeEvolution}) and highly flattened galaxies are inconsistent with heavy mass loss. Our model allows such combinations, but the observations of Crater II disfavour them. Therefore, inclusion of this effect should not significantly affect the results; (ii) we consider a static Milky Way potential -- as the mass of the early Milky Way grows, the pericentre of Crater II will tighten. Our estimates for the proper motion can be considered as upper limits as with a growing potential, we require more radial orbits today to explain the level of total mass loss.

\begin{figure}
$$\includegraphics[width=\columnwidth]{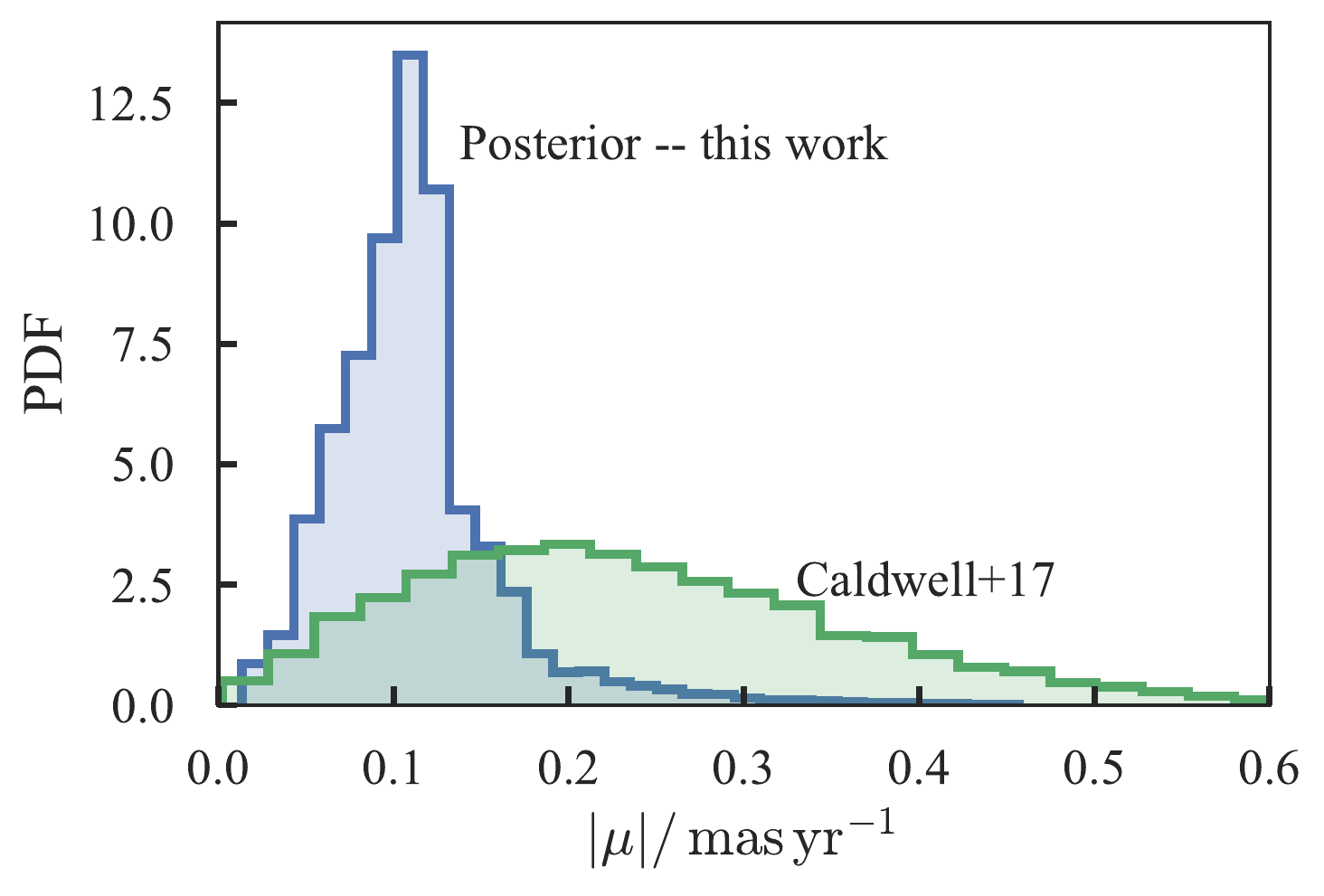}$$
\caption{Posterior distribution on proper motion magnitude in Galactocentric rest frame (blue) alongside the measurement from \protect\cite{Caldwell2017} (green). For the observations of Crater II to be consistent with $\Lambda$CDM we require its observed Galactocentric rest frame proper motion to be less than $0.2\,\mathrm{mas\,yr}^{-1}$.}
\label{Fig::ProperMotion}
\end{figure}

We use the samples on the mass loss to compute the posterior on the proper motion magnitude $|\mu|$ in Figure~\ref{Fig::ProperMotion} (by reweighting the orbit samples shown in Fig.~\ref{Fig::Orbit} by the ratio of the posterior mass loss distribution to the prior mass loss distribution). We find that $|\mu|<0.2\,\mathrm{mas\,yr}^{-1}$ for Crater II to fit into the standard $\Lambda$CDM galaxy formation picture. In equatorial coordinates, this translates to $(\mu_{\alpha*},\mu_\delta)=(-0.21\pm0.09,-0.24\pm0.09)\mathrm{mas\,yr}^{-1}$. The proper motion measurement of \cite{Caldwell2017} is consistent with this interpretation. 

We have run our analysis using the mass loss relation from equation~\eqref{eqn::Mmufit} calibrated to the $s=1$ and $s=0.5$ simulations. This results is very small changes in $1\sigma$ confidence interval on equatorial proper motion to $0.1\,\mathrm{mas\,yr}^{-1}$ and $0.11\,\mathrm{mas\,yr}^{-1}$ respectively. Additionally, we have run our analysis using the flatter tidal tracks from column 3 of Table~\ref{Table::PMN08AlphaBeta}. This results in an increased initial half-light radius for Crater II but the estimate of the proper motion is unchanged. 

Our results also suggest the `accretion' time of Crater II is $(8\pm3)\,\mathrm{Gyr}$, and the initial dynamical-mass-to-light ratio and luminosity were $\log_{10}((M/L)_0/(\mathrm{M}_\odot/\mathrm{L}_\odot))=(1.4\pm0.4)$ and $\log_{10}(L/\mathrm{L}_\odot)=(6.9\pm0.8)$ (we have calibrated these relations using the $s=1$ simulations and give the fitting function parameters in Table~\ref{Table::PMN08AlphaBeta}). This last result seems slightly in tension with the observed metallicity of Crater II of $[\mathrm{Fe}/\mathrm{H}]=-2\,\mathrm{dex}$
\citep{Caldwell2017}, 
whereas the initial luminosity suggests Crater II was comparable to Fornax which has a metallicity of $[\mathrm{Fe}/\mathrm{H}]=-1\,\mathrm{dex}$ \citep{McConnachie2012}. However, Fornax appears to have an extended star formation history with a steady increase in metallicity from $[\mathrm{Fe}/\mathrm{H}]=-1.5\,\mathrm{dex}$ $\sim10\,\mathrm{Gyr}$ ago \citep{deBoer2012}. It is therefore plausible that heavy tidal disruption stifled extended star formation in Crater II suppressing the metallicity.

\section{Weakly Cusped or Cored Dark Haloes}\label{Sec::CoreCusp}
Our choice of a stellar Plummer profile embedded in an NFW halo is somewhat arbitrary. More physically, one might expect that dissipation causes baryons to concentrate more than their host dark matter halo and subsequent star formation will only occur from the densest baryons. Therefore, the initial stellar profile should be more concentrated than the dark matter profile. Subsequent collisionless evolution affects both the light and dark matter equally so a cored dark matter profile would be accompanied by a cored stellar profile. 
These arguments would suggest that the inner stellar profile traces the dark matter (more exactly, our argument suggests that the radius containing a fraction $f$ of the stars is less than or equal to the radius containing $f$ of the dark matter for all $f$). We have found that Crater II could once have had a luminosity of around $10^7\mathrm{M}_\odot$. At these stellar masses, stellar feedback may weaken the central cusp and lead to an altered tidal evolution. For the brighter dSphs like Sculptor or Fornax, modelling of datasets of discrete radial velocities does show strong preference for cored dark halo profiles~\citep[e.g.,][]{Am12,Am13}. However, the APOSTLE simulations~\citep{Sawala2016} manage to reproduce the observed Local Group properties without any stellar feedback model producing cored from dark matter cusps. In this section, we explore how the variations in the stellar and dark profiles could alter our conclusions. 

Weaker cusped models have been explored by \cite{Frings2017}, who find that Crater II can be explained by an initially weaker dark-matter cusp (central logarithmic density slope $\sim0.6$) which produces a broadened stellar profile when placed on a highly eccentric orbit. However, such cored tidal evolution tracks \citep[e.g.][]{Errani2015} appear inconsistent with the observed dSph relationship between dark-matter mass and stellar mass. In the extreme case where mass follows (cored) light, the mass-to-light ratio doesn't evolve whilst the stellar mass decreases. If a cored dSph starts on the mass-to-light ratio versus stellar mass relation followed by all the dSphs, heavy tidal disruption will move it off the relation, whereas disruption of cuspy dSphs produce tracks that align with the observed relation \citep{Penarrubia2008}.

\begin{figure}
$$\includegraphics[width=\columnwidth]{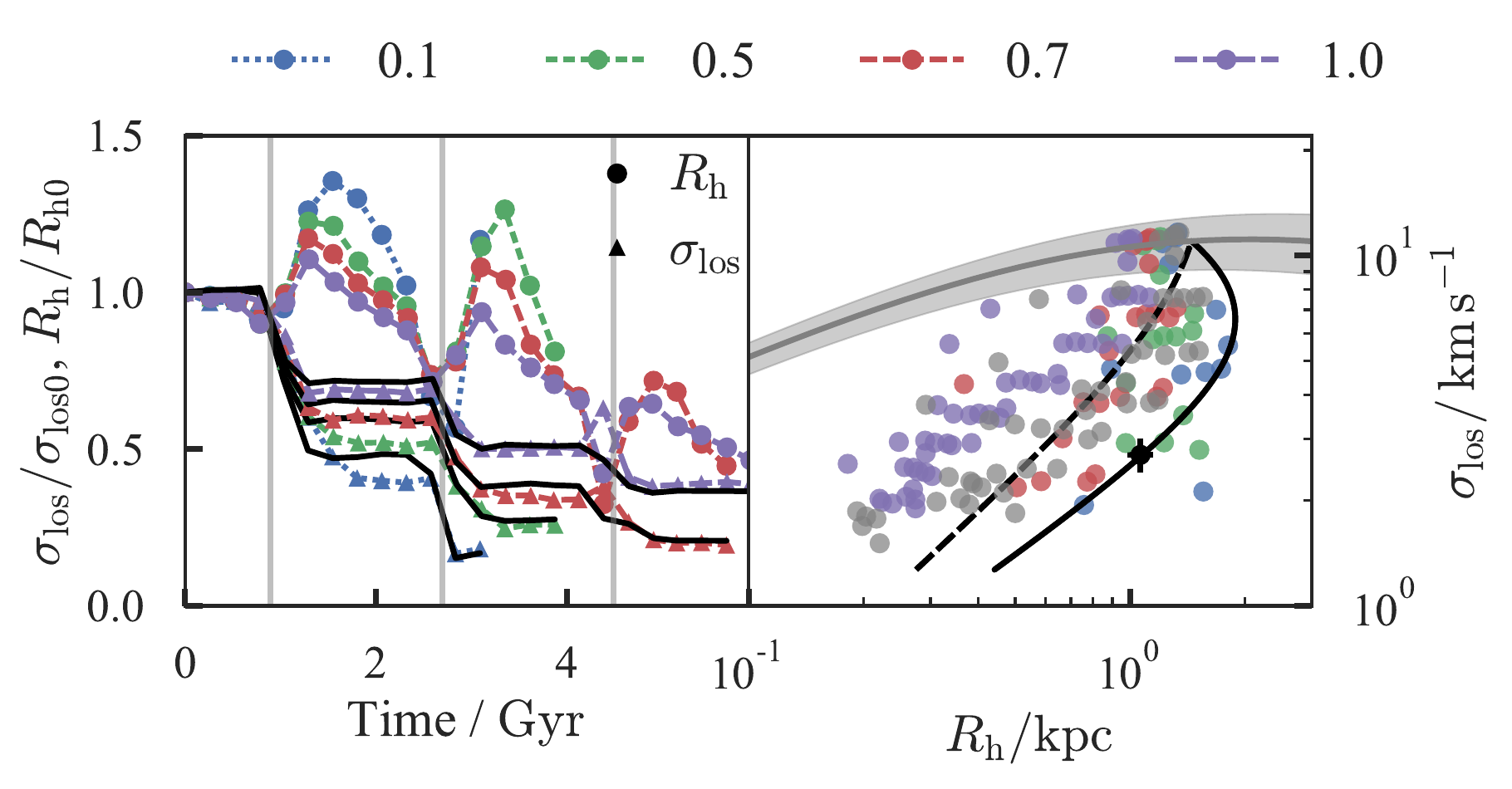}$$
\caption{Observed properties of tidally-disrupted dSphs with varying equal inner stellar ($\gamma_\star$) and dark matter ($\gamma_\mathrm{DM}$) density slope (blue $\gamma=0.1$, green $\gamma=0.5$, red $\gamma=0.7$ and purple $\gamma=1$). The left panel shows the evolution of the projected velocity dispersion (triangles) and half-light radius (circles). Overplotted in black are the predictions for the projected velocity dispersion from the central mass loss. Pericentric passages are shown by grey lines. The right panel is a version of the bottom panel of Fig.~\ref{Fig::SlosRh} with this set of models and the $s=1$ spherical models of Section~\ref{Sec::Nbody} shown in grey.}
\label{Figure::CuspCore}
\end{figure}

We run a further four simulations with $s=1$, $|\mu|=0.1\,\mathrm{mas\,yr}^{-1}$ and $c/a=1$ but varying the inner dark and stellar slopes $\gamma_\mathrm{DM}=\gamma_\star=(0.1,0.5,0.7,1)$. We set the mass within $1.3r_\star$ (approximately the 3D half-mass radius) to be equal to that for the Plummer embedded in NFW models of Section~\ref{Sec::Nbody} and recompute the initial tidal radius. We show the results of the time evolution of the projected velocity dispersion and the half-light radius for the models in Fig.~\ref{Figure::CuspCore}. We find that the central velocity dispersion is insensitive to details of the stellar and dark profiles and solely a function of the central mass loss. As in Fig.~\ref{Fig::SlosRh}, the evolution of the half-light radius is a more complicated function of the profile details. We find that the $\gamma_\mathrm{DM}=1$ simulations mirror the tracks traced by our original simulations (modulo a small difference in initial half-light radius) leading us to conclude that the details of the stellar profile are unimportant. We find that more cored dark-matter profiles produce a ballooning of the stellar radius after pericentric passage. This is possibly because, for a fixed energy injection, stars are allowed to stray further in a cored potential. This results in the more cored galaxies following a steeper path in the $\sigma_\mathrm{los}$ against $R_\mathrm{h}$ plane. However, naturally cored profiles are less resilient to the tidal field such that the simulation with $\gamma_\mathrm{DM}=0.1$ only lasts two pericentric passages. This though may alleviate some of the tension of our previous simulations with Crater II. If Crater II was initially a weakly-embedded system with a weaker dark matter cusp, tidal disruption will primarily act to reduce the velocity dispersion, whilst still retaining a large half-light radius. In this picture however, Crater II has only recently been accreted, meaning that it should be more closely associated with its shed material and potentially other systems if part of a group infall.

In conclusion, we find that variations in the stellar profile within a fixed dark matter potential do not alter the conclusions of the previous section. However, cored dark matter profiles produce a significantly steeper tidal track. For cored profiles, we still require significant ($\sim90\percent$) mass loss to explain Crater II's velocity dispersion, but the associated half-light radius is more consistent with that of Crater II.

\section{Conclusions}\label{Sec::Conclusions}

We have explored routes for dSphs to possess suppressed velocity dispersion, with a particular application to the recently-discovered Crater II\footnote{The code and notebooks used for analysing the simulations and generating the figures in this paper are made available at \url{https://github.com/jls713/CraterII}.}. Our conclusions are as follows:

\begin{enumerate}
\item We have confirmed the tidal evolutionary tracks of \cite{Penarrubia2008} for cusped dark matter profiles and extended their applicability to flattened (axis ratios $0.3$) and weakly embedded systems ($r_{s\mathrm{DM}}=r_{s\star}, 0.5r_{s\star}$). We find that the weakly-embedded and flattened systems follow the expected tidal tracks in velocity dispersion. However, weakly-embedded systems produce slightly different tracks of half-light radius with mass loss. The half-light radius is significantly suppressed, producing a flatter track in the velocity-dispersion vs. half-light radius plane (irrespective of the details of the stellar profile). This is problematic for the observations of Crater II which, within standard galaxy formation theory, is expected to be a weakly-embedded system but is only marginally consistent with the cusped simulations. This tension is eased if the central dark matter cusp of Crater II is weakened. More cored dark matter profiles follow similar tidal tracks in velocity dispersion but their half-light radii can increase between pericentric passages with more cored profiles showing a larger increase.
\item We have demonstrated that there are two competing effects driving the shape evolution of flattened dwarf galaxies. Tidal shocking causes the central regions to become rounder whilst tidal distortion generates a prolate tidally-locked envelope. Initially spherical systems are driven to gradually weakly prolate/triaxial structures through tidal distortion whilst initially flattened systems are tidally shocked to rounder distributions before heavy total mass loss causes tidal distortion to effect the inner regions. The central axis ratios are solely functions of the initial shape central total mass loss (irrespective of orbital history) and the flattened and spherical evolutionary sequences converge after $\sim70\percent$ total mass loss. At late times, a heavily tidally-disrupted system is approximately prolate-triaxial (axis ratios 0.8 and 0.6) independent of the initial shape.
\item Although both shape and variation in the `universal' halo aid in explaining Crater II's velocity dispersion, the dominant effect has to be significant tidal stripping. We estimate Crater II has lost $\sim90\percent$ of its initial central mass and as such needs to be on a reasonably eccentric orbit for multiple pericentric passages. The proper motion of Crater II today must be $(\mu_{\alpha*},\mu_\delta)=(-0.21\pm0.09,-0.24\pm0.09)\mathrm{mas\,yr}^{-1}$ for the results to fit into standard galaxy formation theory. 
\end{enumerate}

\subsection{Observational Prospects}

We close by discussing briefly the observational tests that our study suggests. A measurement of the line-of-sight depth of Crater II would be informative for two reasons: (i) we have shown that highly oblate systems can have suppressed velocity dispersions so we would like to assess the importance of this effect, (ii) shape correlates with tidal evolution so a highly flattened system suggests little tidal evolution. Currently, no dSphs have line-of-sight depth measurements as these are very challenging observations. However, as Crater II is a large system it is perhaps a good candidate for such a measurement (although it has few stars). Assuming Crater II is spherical, we expect a magnitude width of $\Delta M = (5/\ln10) R_\mathrm{h}/D = 0.02\,\mathrm{mag}$. We might optimistically be able to measure the width of the horizontal branch to this degree of accuracy although we require highly reliable models \citep{Gratton2010}.

The best test of the theory presented here is a robust measurement of Crater II's proper motion using astrometric data from the Gaia mission \citep{Prusti2016}. Crater II has a number of giant stars, $(g-r)=0.8$, at $r=18.4$. Using the final Gaia dataset, it should be possible to measure the proper motions of these stars accurate to $0.08\masyr$ (computed using the \texttt{pygaia} package). With proper motion measurements of multiple member stars, it will be possible to rule out heavy tidal disruption if the mean proper motion $|\mu|>0.2\masyr$. However, if $|\mu|<0.2\masyr$, this would confirm tidal disruption as a likely cause of the velocity dispersion suppression of Crater II .

A further useful observation is the tangential velocity (or proper motion) dispersion. Assuming sphericity, Crater II should have a proper motion dispersion of $5\,\mu\mathrm{as\,yr}^{-1}$. A significantly oblate figure (e.g. axis ratio 0.3) will have a larger tangential dispersion (by $\sim68\percent$). Such an accurate observation seems beyond the reach of Gaia. However, longer baseline observations may probe down to this accuracy. For instance, \cite{Massari2017} have recently used the $\sim12$ year baseline between Hubble Space Telescope and Gaia observations of 15 stars in Sculptor (significantly brighter than Crater II, $L\sim3\times10^6\mathrm{L}_\odot$ but at a similar distance $84\,\mathrm{kpc}$) to measure the tangential dispersions accurate to $\sim4\,\mathrm{km\,s}^{-1}$. The internal velocity dispersion can be measured significantly more accurately than the mean tangential velocity as it is insensitive to uncertainties in the zero-point.  

Finally, with proper motion filtering using Gaia, it will be possible to search for extra-tidal material for many dSphs. The prediction here is that Crater II has lost $\sim70\percent$ of its stellar component, so we anticipate a significant associated tidal stream. If Crater II has a cored dark matter profile, the stellar material must have been stripped within the last one or two pericentric passages so should be more closely spatially associated with Crater II. For a cusped dark matter profile, the material could have been stripped over multiple pericentric passages and will be more dispersed making its detection more difficult.

\section*{Acknowledgements}
JLS acknowledges the support of the STFC. We have made use of the \texttt{corner} package from \cite{ForemanMackey2014} and the \texttt{pygaia} package kindly provided by the Gaia Project Scientist Support Team and the Gaia Data Processing and Analysis Consortium (DPAC). We thank useful comments and conversations with members of the Cambridge streams group.




\bibliographystyle{mnras}
\bibliography{bibliography.bib} 




\bsp	
\label{lastpage}
\end{document}